\documentclass[twocolumn,showkeywords,preprintnumbers,amsmath,aps]{revtex4-1}
\usepackage{graphicx}
\usepackage{dcolumn}
\usepackage{bm}
\usepackage{color}
\usepackage{soul}
\usepackage{microtype}
\usepackage{amssymb}

\begin{document}
\title{Gap states and valley-spin filtering in transition metal dichalcogenide monolayers}

\author{Dominik Szcz{\c{e}}{\'s}niak$^{1, 2}$}\email{dszczesn@purdue.edu; d.szczesniak@ujd.edu.pl}
\author{Sabre Kais$^{1}$}


\affiliation{$^1$Department of Chemistry, Purdue University, 560 Oval Dr., 47907 West Lafayette, Indiana, United States of America}
\affiliation{$^2$ Department of Theoretical Physics, Faculty of Science and Technology, Jan D{\l}ugosz University in Cz{\c{e}}stochowa, 13/15 Armii Krajowej Ave., 42200 Cz{\c{e}}stochowa, Poland}

\date{\today} 
\begin{abstract}

The magnetically-induced valley-spin filtering in transition metal dichalcogenide monolayers ($MX_{2}$, where $M$=Mo, W and $X$=S, Se, Te) promises new paradigm in information processing. However, the detailed understanding of this effect is still limited, regarding its underlying transport processes. Herein, it is suggested that the filtering mechanism can be greately elucidated by the concept of metal-induced gap states (MIGS), appearing in the electrode-terminated $MX_{2}$ materials {\it i.e.} the referential filter setup. In particular, the gap states are predicted here to mediate valley- and spin-resolved charge transport near the ideal electrode/$MX_{2}$ interface, and therefore to initiate filtering. It is also argued that the role of MIGS increases when the channel length is diminished, as they begin to govern the overall valley-spin transport in the tunneling regime. In what follows, the presented study yields fundamental scaling trends for the valley-spin selectivity with respect to the intrinsic physics of the filter materials. As a result, it facilitates insight into the analyzed effects and provide design guidelines toward efficient valley-spin filter devices, that base on the discussed materials or other hexagonal monolayers with a broken inversion symmetry.

\end{abstract}
\maketitle
%

\section{Introduction}

Most of the present concepts behind the electronic control of information rely on the manipulation of charge flow or spin angular momentum of electrons. However, recent developments in quantum electronics show that it is also possible to address an alternative property of the electron, namely its valley pseudospin \cite{gunawan, rycerz, xiao1, mak, aivazian, ye1}. In comparison to its charge and spin counterparts, the valley degree of freedom constitute binary index for the low-energy electrons associated with the local conduction band minima (valleys) in the momentum space of a crystal \cite{gunawan}. As a result, it is expected that the valley-based (valleytronic) devices should provide new or improved functionalities in the field of classical and quantum information processing {\it e.g.} in terms of low-power valley or hybrid valley-spin logic devices \cite{ang, schaibley, vitale} as well as complex qubit basis sets \cite{rohling, vitale}. Nonetheless, to efficiently perform valleytronic operations in solid state systems, it is required to have control over the selective population of distinguishable valleys, toward their polarization \cite{xiao1, mak, aivazian, ye1}. Moreover, the electrons should occupy polarized valleys long enough to allow logic operations of interest \cite{schaibley, vitale}.

Given the above background, not all solid state materials that exhibit local energy extrema in the momentum space are well suited for the valley control of information. From among the systems already considered as potential hosts for valleytronics \cite{takashina, gunawan, rycerz, zhu, isberg}, the most promising are the two-dimensional (2D) layered crystals with honeycomb structures, due to their strong valley-selective coupling with the external fields \cite{schaibley, vitale}. In the family of such 2D systems, currently of particular attention are the group-VIB transition metal dichalcogenide monolayers ($MX_{2}$, where $M$=Mo, W and $X$=S, Se, Te) \cite{schaibley}. Similar to the graphene, the $MX_{2}$ materials possess two inequivalent but energetically degenerate valleys at the $K$ and $K'$ high symmetry points in their first hexagonal Brillouin zone \cite{chhowalla}. However, the $MX_{2}$ monolayers are also characterized by the inherently broken inversion symmetry and the strong spin-orbit coupling (SOC). In what follows, they exhibit direct semiconducting band gaps and can benefit from the chiral optical selection rules toward dynamical control of the valley population \cite{mak, ye2}. The same properties lead also to the coupling between the valley and spin degrees of freedom (the valley-spin locking) \cite{xiao2} and allow control of their polarization by the means of the out-of-plane external magnetic field \cite{aivazian, macneill} or the magnetic exchange field \cite{qi, zhao}.

In terms of the information processing in $MX_{2}$ materials, the idea to manipulate valley pseudospin via the magnetic field effects appears so far to be more robust than the use of the optical pumping methods \cite{xu, hsieh}. In particular, recent studies show that the strong valley-spin splitting can be practically obtained via magnetic exchange fields, that allow to effectively overcome issue of large magnetic fields required for polarization \cite{qi, zhao, xu, norden}. Simultaneously, solutions developed within magnetic approaches clearly inspire early attempts in electrical generation and control of valley carriers in selected $MX_{2}$ monolayers \cite{ye1, hsieh, hung, gut}. Most importantly, however, the discussed method of control allows to utilize the $MX_{2}$ materials as a magnetic channel contacts between metallic leads (the so-called two-terminal setup) to perform valley- and spin-resolved switching operations, by the analogy to the well-established concept of the spin-filter \cite{appelbaum, moodera}. The described valley-spin filter received already notable consideration, initially in terms of the graphene-based systems \cite{rycerz, xiao1, firozislam, huang, thompson} and later based on the discussed here $MX_{2}$ monolayers \cite{yuan, majidi, rostami, tahir}. However, although mentioned above studies provide successful initial modeling of the $MX_{2}$ valley-spin filters, the in-depth discussion of the underlying transport phenomena is still absent in the literature, hampering further developments in the field.

In this context, the present study attempts to provide new contribution to the primary understanding of the valley-spin filtering effect in $MX_{2}$ materials. In particular, it is proposed here that the filtering mechanism can be largely elucidated within the concept of metal-induced gap states, that appear in the electrode-terminated $MX_{2}$ monolayers (the referential filter setup). Such insight is meant to include, usually ignored, inherent and distinct electronic features of the $MX_{2}$ materials {\it i.e.} their multiband structure with complex orbital symmetry behavior, the strong spin-orbit coupling, as well as the Berry curvatures. As a result, this analysis intends to not only facilitate the fundamental understanding of the discussed processes but also to provide their general trends with respect to the pivotal control parameters such as the magnetic field strength, transport channel length and Fermi level position in the semiconductor. Hence, the results are expected to be of importance for the future design of valley-spin filter devices for information processing, as build by using the $MX_{2}$ monolayers or potentially other 2D hexagonal materials with broken inversion symmetry.

\section{Theoretical Model}

To understand mechanism of the valley-spin filtering in $MX_{2}$ materials, this effect is addressed here within the referential two-terminal filter setup. Therein, the $MX_{2}$ monolayer of interest is terminated by the metal electrodes and exposed to a magnetic field that induces valley and spin polarization. In what follows, the $MX_{2}$ monolayer acts as a transport channel that allows to achieve different transmission probabilities for a charge carriers with opposite valley or spin degrees of freedom. According to the available studies, it is important to note here that such $MX_{2}$ channel is expected to be relatively short ($\leqslant 10$ nm) \cite{yuan, hsieh, majidi}. Thus, the large scattering of charge carriers can be avoided toward their better mobility \cite{liu1} as well as improved valley coherence and lifetime \cite{vitale}. However, the diminished geometry of the $MX_{2}$ channel results also in its strong hybridisation with the attached metal contacts. As an outcome, the electrodes cause locally metallic character of the $MX_{2}$ material in the vicinity of the interface \cite{kerelsky} or even lead to the entirely metallic behavior of a short $MX_{2}$ channels \cite{wang}.

In the present study, it is argued that the mentioned electronic nature of the electrode-terminated $MX_{2}$ materials is likely to have significant impact on the valley-spin filtering effect. Of particular importance in this regard are the peculiar metal-induced gap states (MIGS), responsible for the locally metallic character of the semiconductor near the metal-semiconductor interface. By following Heine \cite{heine} and Tersoff \cite{tersoff} in spirit, at the ideal bulk metal-semiconductor junction the MIGS are formed from the propagating states of the metal that extend and decay into the forbidden energy region of the semiconductor. Recent studies confirm that such states exist also at the low-dimensional metal/$MX_{2}$ junctions and largely determine their interfacial physics \cite{sotthewes, kerelsky, szczesniak1, szczesniak2, guo}. One of the most important aspects of these states is their role in mediating the Fermi level ($E_{F}$) pinning effect at the metal/$MX_{2}$ interface \cite{sotthewes, szczesniak2, guo}. This is to say, the MIGS define charge neutrality level near the interface and therefore largely control charge transport processes across this region \cite{sotthewes, rai, szczesniak1, szczesniak2}. In what follows, when the valley and spin polarization is induced in the $MX_{2}$ filter, the charge injection governed by the MIGS is likely to become polarized as well. This observation is additionally reinforced by the fact that MIGS constitute inherent property of a semiconductor \cite{tersoff, reuter} and should respond to the external fields in a similar way as their bulk counterparts. In this regard, the argued role of the MIGS in the valley-spin filtering may increase even further when the $MX_{2}$ monolayer length is diminished and the corresponding transport processes across the channel approach tunneling regime. In particular, the MIGS that couple to the metallic states at the $E_{F}$ and exhibit small decay rates are able to penetrate semiconducting band gap deep into the channel \cite{heine, kane}. In a short channel limit such states survives long enough to mediate transport across the entire tunneling gap \cite{mavropoulos, heine, kane}. Having all above in mind, the MIGS may provide important insight into the primary mechanism of the valley-spin filtering effect in $MX_{2}$ monolayers.

To verify the role of MIGS in the discussed spin-valley filtering effect, it is instructive to analyze their behavior in the momentum space. This can be done by solving the inverse eigenvalue problem (IEP) \cite{reuter, szczesniak1}, assuming that wavevector ($\mathbf{k}$) in a solid takes on complex values. The matrix form of the IEP can be written as:
\begin{equation}
\label{eq1}
\left(\mathbf{H}_{i}-\mathbf{H}_{i+1}\vartheta-...-\mathbf{H}_{i+j-1}\vartheta^{j-1}-\mathbf{H}_{i+j}\vartheta^{j}\right) \Psi=0,
\end{equation}
where $\mathbf{H}_{i}$ and $\mathbf{H}_{i+j'}$ denote component Hamiltonian matrices for the origin $i^{th}$ unit cell and its interactions with the neighboring cells, respectively. Moreover in Eq. (\ref{eq1}), $\vartheta=e^{{\rm i} \mathbf{k} \mathbf{R}}$ is the generalized Bloch phase factor for a given $\mathbf{R}$ lattice vector, while $\Psi$ stands for the wave function column vector. In case of the $MX_{2}$ materials, the lattice vector takes form $\mathbf{R} = \alpha\mathbf{a}_{x}+\beta\mathbf{a}_{y}$, where $\mathbf{a}_{x}$ and $\mathbf{a}_{y}$ describe the primitive vectors of 2D hexagonal lattice, whereas $\alpha$ and $\beta$ are integer values. According to that, $j'\in \left[ 1, 2, \dots, j-1, j \right]$, where $j=\alpha$ or $\beta$, depending on the chosen crystal direction.

To account for all the important electronic properties of the $MX_{2}$ crystals, herein these materials are described within the following magnetized Hamiltonian:
\begin{equation}
\label{eq2}
\mathbf{H} = \mathbf{H}_{TB} + \mathbf{H}_{\rm SOC} + \mathbf{H}_{\rm B}.
\end{equation}
In Eq. (\ref{eq2}), the $\mathbf{H}_{TB}$ part stands for the $6 \times 6$ tight-binding Hamiltonian, which includes up to the third-nearest-neighbor interactions and is constructed within the \textls[-18]{$\{ \left| d_{z^2}, \uparrow \right>, \left| d_{xy}, \uparrow \right>, \left| d_{x^2-y^2}, \uparrow \right>, \left| d_{z^2},\downarrow \right>, \left| d_{xy}, \downarrow \right>, \left| d_{x^2-y^2}, \downarrow \right> \}$} minimal basis for the $M$-type atoms, as shown in \cite{liu2}. Next, the $\mathbf{H}_{\rm SOC} = \lambda \mathbf{L} \cdot \mathbf{S}$ is the intra-atomic SOC term, in which $\lambda$ gives the spin-orbit coupling constant, whereas $\mathbf{L}$ and $\mathbf{S}$ are the orbital and spin angular momentum operators, respectively. Finally, the $\mathbf{H}_{\rm B}=-\mu\sigma_{\rm z} \otimes \mathbf{I}_{3 \times 3}$ describes the influence of the external magnetic field, which is perpendicular to the $MX_{2}$ plane. Therein, $\mu = g \mu_{B} \mathbf{B}$ is the Zeeman energy, where $g=2$ is the gyromagnetic factor for the $d$-type orbitals, $\mu_{B}$ stands for the Bohr magneton, and $\mathbf{B}$ is the magnetic field in Teslas. Moreover, in $\mathbf{H}_{\rm B}$, the $\sigma_{\rm z}$ describes Pauli spin matrix, $\otimes$ is the Kronecker product and $\mathbf{I}_{3 \times 3}$ stands for the $3 \times 3$ identity matrix. The tight-binding and $\lambda$ parameters are adopted from \cite{liu2}.

Due to the specific symmetry-based character of the Hamiltonian (\ref{eq2}), the IEP of Eq. (\ref{eq1}) retains its general nonlinear form with respect to $\vartheta$, and has to be solved by the linearization methods \cite{szczesniak2}. As a results, the IEP yields the pairs of spin-dependent $\vartheta$ and $1/\vartheta$ solutions, which are linked by the time-reversal symmetry for each of the distinct valleys at the $K$ and $K'$ high symmetry points, respectively. In this manner, the eigenvalues of IEP corresponds to the propagating states when $|\vartheta|=1$ and to the decaying states when $|\vartheta|<1$. The combination of such IEP solutions is referred here to as the complex band structure (CBS), employed to directly relate the filtering processes of interest to the intrinsic electronic properties of the $MX_{2}$ monolayers.

\section{Numerical results}

\begin{figure}[ht!]
\includegraphics[width=\columnwidth]{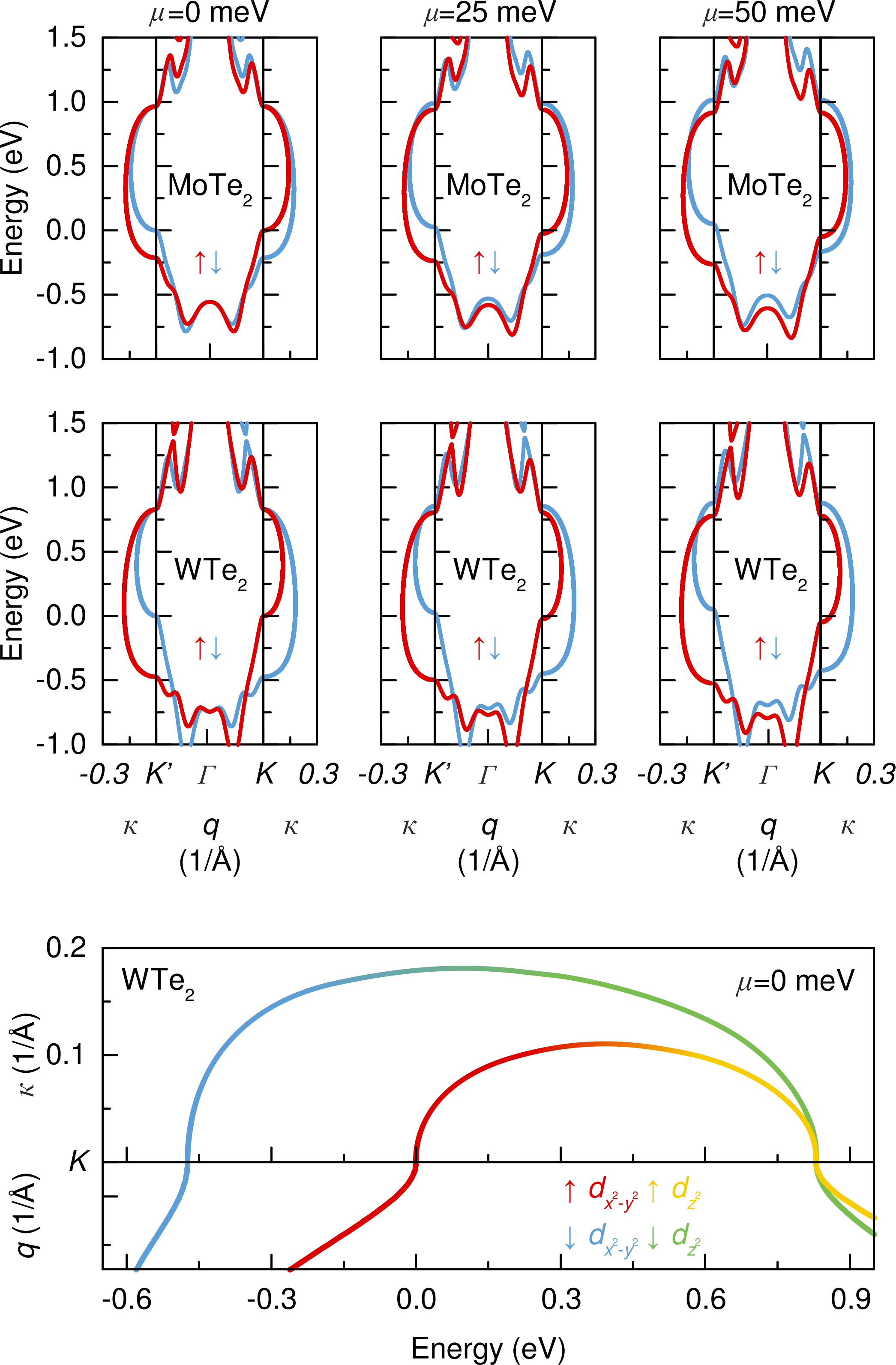}
\caption{The valley- and spin-resolved complex band structures of the MoTe$_2$ (first row) and WTe$_2$ (second row) monolayers for different values of the out-of-plane magnetic field (columns). The zero reference energy is set at the top-edge of the valence band, and the momentum axis is given in the unified unit of $1/$\AA. The orbital-projected complex band structure in pristine WTe$_2$ near the $K$ point (lowest panel).}
\label{fig01}
\end{figure}

In Fig. \ref{fig01}, the CBS solutions under selected values of the out-of-plane magnetic field are depicted for the representative MoTe$_2$ (first row) and WTe$_2$ (second row) materials. Note that these materials exhibit the strongest valley- and spin-related effects among the Mo- and W-based monolayers, respectively. The middle panel of each sub-figure presents the spin-dependent propagating states for $q=$Re$[\mathbf{k}]$ along the $K'-{\it \Gamma}-K$ path in the hexagonal Brillouin zone. These states capture the band-edge properties in the vicinity of the band gap {\it e.g.} the large spin splitting of the valence band due to the SOC. On the other hand, the left and right panel of each subfigure presents the respective spin-dependent gap states for $\kappa=$Im$[\mathbf{k}]$ at the $K'$ and $K$ points. Herein, only the gap solutions with the smallest $\kappa$ for each spin orientation are presented, by arguing the fact that they describe the most penetrating states within the gap. This is according to the corresponding decay of the wave function per unit cell as given by $e^{-{\rm Im}[\kappa]a}$, where $a$ is the lattice constant. Therefore, these are the states described by the the lowest decay rates ($\kappa$), or the longest decay lengths ($1/\kappa$), that can be interpreted as the MIGS and suppose to provide the major contribution to the transport processes of interest.

In this context, the first observation arising from the presented results corresponds to the analytical character of the depicted decaying states, that visibly inherit spin properties of their propagating counterparts. Although less obvious, the discussed states also follow the orbital character of the propagating solutions. In particular, the orbital character of decaying states changes from the majority $d_{x^2-y^2}$-type behavior in the vicinity of the donor-like valence band into the more $d_{z^2}$-type symmetry close to the acceptor-like conduction band. Please see the lowest panel of Fig. \ref{fig01} for the graphical representation of the orbital-projected CBS in the representative pristine WTe$_2$ monolayer. To present the transition of the orbital character in the most transparent way, the CBS is plotted near the $K$ point. As shown therein, the change in the described orbital character occurs smoothly as the energy approaches the branch point of the semielliptic decaying state. This finding is in agreement with the fact that evanescent states within the gap constitute analytical continuation of the corresponding real bands \cite{reuter, szczesniak1, tersoff}. Note that the same behavior is observed in other $MX_{2}$ materials considered in the present study. In what follows, the decaying states clearly constitute the inherent property of the semiconductor and the direct continuum of the propagating states within the gap, in agreement with the MIGS character. For more information on the orbital-projected character of the real band structures in the $MX_{2}$ materials, as described by the employed here $\mathbf{H}_{TB}$ Hamiltonian, please refer to \cite{liu2}.

The results presented in the first and second row of Fig. \ref{fig01} allow also to trace changes of the electronic behavior with respect to the applied out-of-plane magnetic field. In particular, the propagating and decaying states respond in a conventional way to the Zeeman effect induced by the external magnetic field {\it i.e.} states with spins parallel to the field are lowered and those antiparallel raised in energies. Moreover, the relation between the total band gap values at the $K$ and $K'$ points is in good agreement with the experimental exciton charge measurements under out-of-plane magnetic field \cite{macneill}. Yet, in terms of the filtering effect, the most important observation is the relative shift in energies between the spin-dependent decaying states in the $K$ and $K'$ valleys, when the magnetic field takes on nonzero value. Since these states become populated up to the Fermi level near the electrode/$MX_{2}$ interface \cite{sotthewes, kerelsky, szczesniak2, guo}, the observed polarization is expected here to initiate valley- and spin-selectivity of the charge transport in this region. This is to say, the electrode-injected charge is predicted to be valley- and spin-filtered via MIGS, before it is transmitted into the bulk propagating states of the $MX_{2}$ channel. This observation is reinforced by the fact that MIGS are the dominant metallic states at the $E_{F}$ near the interface \cite{tersoff, kerelsky, szczesniak2}, and according to Fermi-Dirac distribution, they play central role in the corresponding charge transport \cite{datta}. Moreover, as already shown in the present study, the MIGS directly connect to the valley- and spin-dependent propagating states, allowing selective charge injection. Note that the above argumentation is of particular importance for the ideal electrode/$MX_{2}$ interface where charge injection occurs mainly due to the field emission process {\it i.e.} charge from the electrode is injected into the semiconducting channel via tunneling across MIGS at the $E_{F}$. Nonetheless, even in the case of the phonon-assisted injection the MIGS should not be neglected, as they span the entirety of a band gap near the interface. This aspect is however beyond present study and should be investigated further in the framework of more sophisticated models.

Following the above findings, it is next important to discuss efficiency of the valley- and spin-filtering effect via MIGS, as given by the employed theoretical model. In this respect, one should note that the MIGS polarization is likely to change not only with the magnetic field strength but also as a function of the Fermi level position and the allowed MIGS decay distance. Note also that the last parameter is determined by the length of the region over which MIGS are allowed to decay into the semiconducting gap. In this regard it is important to observe that for a short channels, the MIGS are expected here to mediate transport across the entire $MX_{2}$ material, in addition to the electrode/$MX_{2}$ interface region. This is due to the fact that when the length of the $MX_{2}$ region becomes comparable to the characteristic decay length of a given gap state, charge can tunnel across the channel \cite{mavropoulos, heine, kane}.

\begin{figure*}[ht!]
\includegraphics[width=0.98\columnwidth]{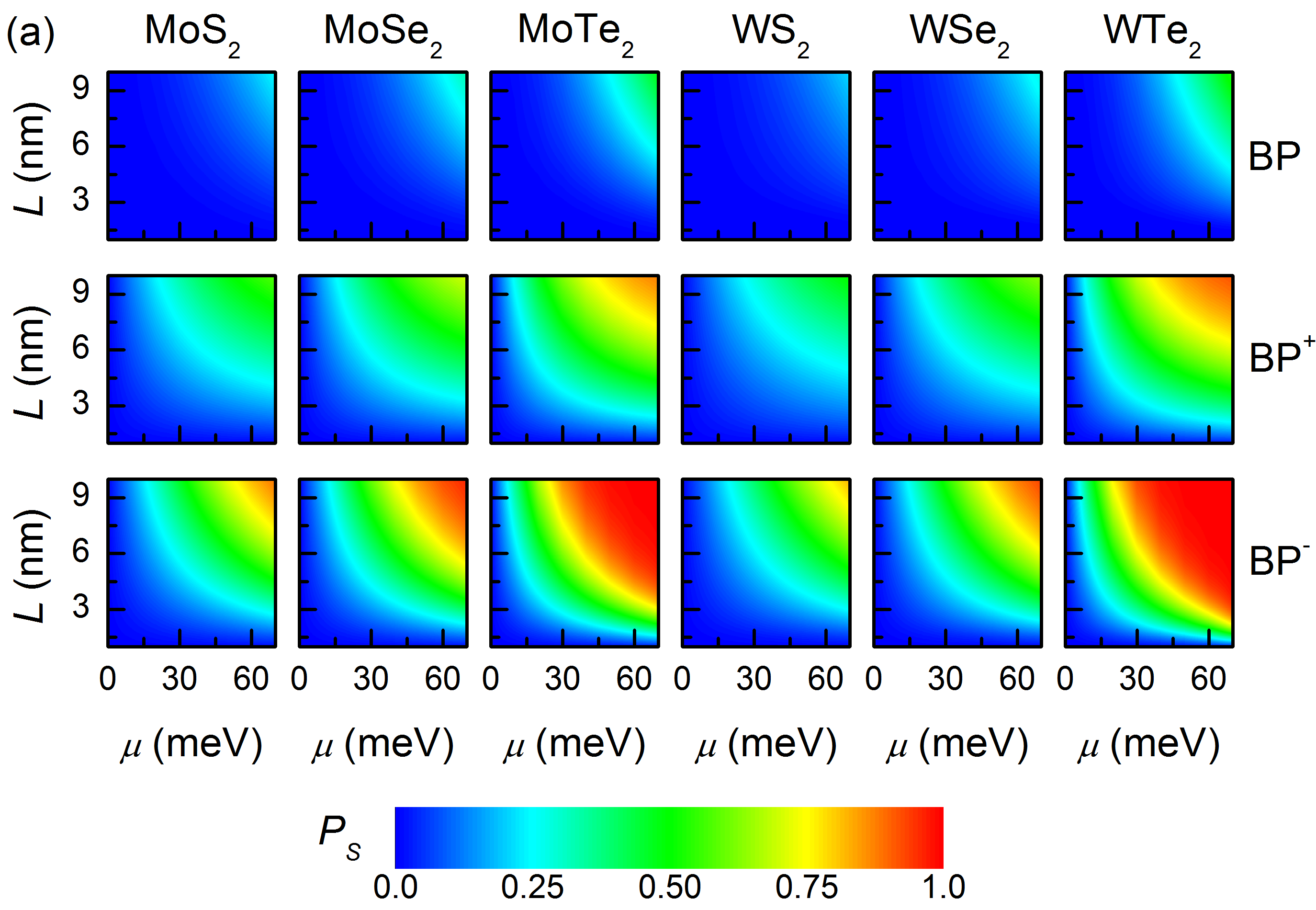}\hspace{0.04\columnwidth}\includegraphics[width=0.98\columnwidth]{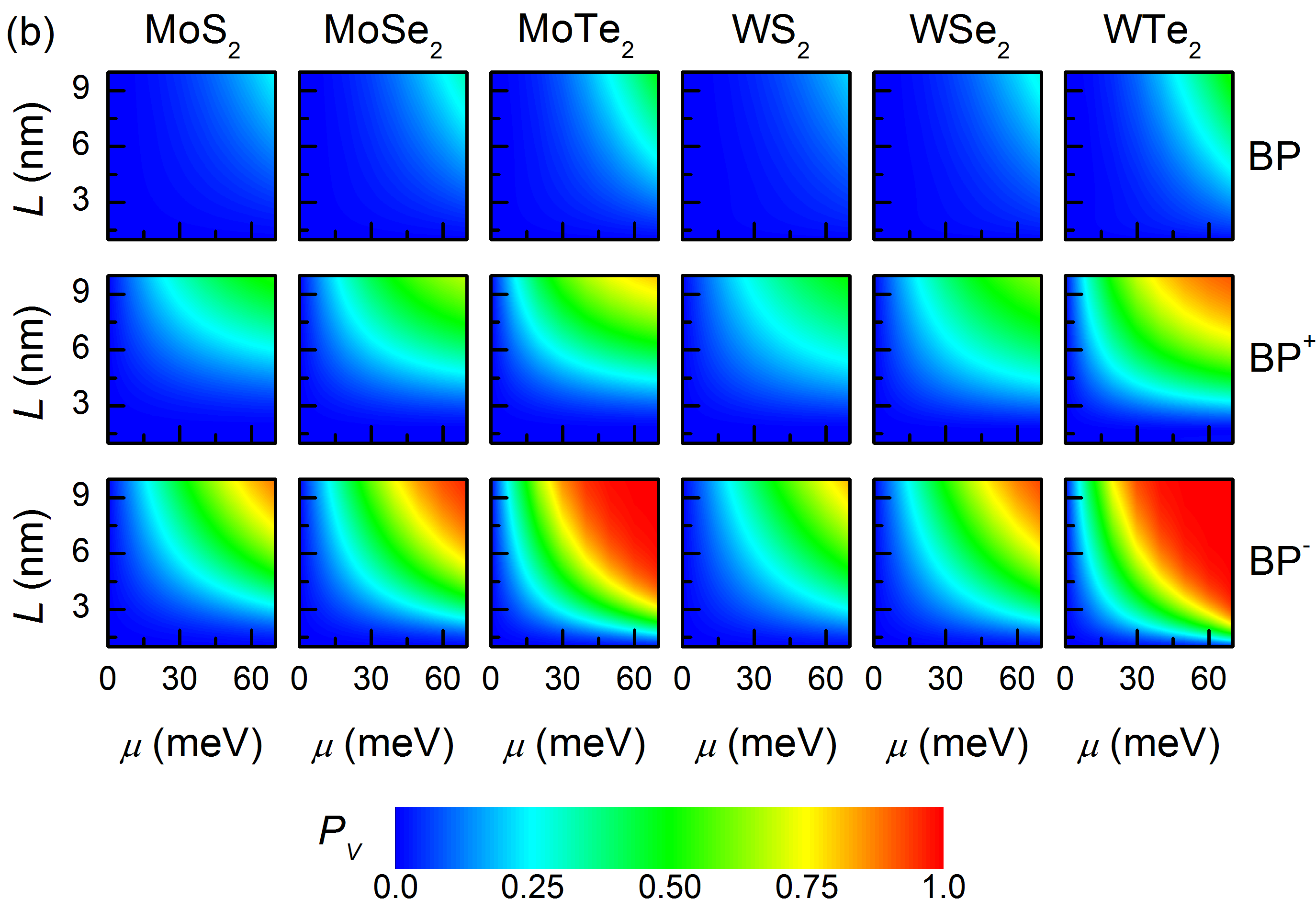}
\caption{The total spin (a) and valley (b) polarization of the tunneling probability decay in $MX_{2}$ monolayers as a function of the external magnetic field strength ($\mu$) and the decay region length ($L$). The results are depicted for three different positions of the Fermi level within the semiconducting band gap, from top to down for the BP, BP$^+$, and BP$^-$, respectively.}
\label{fig02}
\end{figure*}

To investigate influence of the mentioned parameters on the filtering effect, it is first instructive to directly relate MIGS to the tunneling probabilities. In reference to the mentioned wavefunction decay characteristics, the valley- and spin-dependent decay of the tunneling probabilities in the $MX_{2}$-based filters can be given as $T_{K/K', \uparrow/\downarrow}=e^{-2{\rm Im}[\kappa_{K/K', \uparrow/\downarrow}]L}$, where $L$ is the length of the decay region. Note, that the tunneling probability decay should be calculated at the Fermi level to account for the dominant current contributions in the gap region. Herein, this level is not known {\it a priori}, since the metallic contacts are not included explicitly in the present analysis. However, it is possible to set the canonical position of the Fermi level at the branch point of the decaying state, that is characterized by the lowest $\kappa$ value. Note that this approximation refers to the results presented previously for the Fermi level pinning phenomena at the metal-$MX_{2}$ junctions \cite{szczesniak2}. According to that, in the present study the reference level for the calculations is associated with the branch point of the spin-up MIGS in the $K$-valley (BP). Moreover, to account for the moderate Fermi level engineering two additional positions are considered, namely BP$^{+}$=BP+0.25 eV and BP$^{-}$=BP-0.25 eV. In what follows, it is possible now to investigate the total spin polarization:
\begin{eqnarray}
\label{eq3}
P_{S}&\equiv&\frac{T_{K, \uparrow}-T_{K, \downarrow}+T_{K', \uparrow}-T_{K', \downarrow}}{T_{K, \uparrow}+T_{K, \downarrow}+T_{K', \uparrow}+T_{K', \downarrow}},
\end{eqnarray}
as well as the corresponding total valley polarization:
\begin{eqnarray}
\label{eq4}
P_{V}&\equiv&\frac{T_{K, \uparrow}+T_{K, \downarrow}-T_{K', \uparrow}-T_{K', \downarrow}}{T_{K, \uparrow}+T_{K, \downarrow}+T_{K', \uparrow}+T_{K', \downarrow}}.
\end{eqnarray}
In Fig. \ref{fig02}, the total spin and valley polarizations of the tunneling current decay in $MX_{2}$ monolayers are presented as a function of the out-of-plane magnetic field strength and the semiconducting channel length. The first row of subfigures corresponds to the solutions at the BP level, whereas next two rows refer to the results at the BP$^{+}$ and BP$^{-}$ energies, respectively. Moreover, to cover conventional spatial sizes of the $MX_{2}$ channel in the filter setup, it is assumed that $L \in \left[ 1, 10 \right]$ nm. On the other hand $\mu \in \left[ 0, 70 \right]$ meV, so that the BP$^{+}$ and BP$^{-}$ levels are always within the energy gap range.

In general, the results presented in Fig. \ref{fig02} show that the decay of tunneling currents in $MX_{2}$ crystals exhibits similar trends in terms of the valley and spin selectivity. In particular, the polarization of tunneling currents increases along with $\mu$ and $L$, although it can be induced only by the nonzero magnetic field. The observed growth of the $P_{S}$ and $P_{V}$ with the magnetic field strength can be associated with the Zeeman effect, by recalling described before behavior of the MIGS. On the other hand, the $P_{S}$ and $P_{V}$ increases as a function of $L$, since higher value of $L$ provides bigger contribution to the exponential form of $T_{K/K', \uparrow/\downarrow}$. Note, however, that the tunneling current is expected to be particularly effective up to $L \sim 2-3$ nm (depending on the $MX_{2}$ monolayer), due to the maximum decay length of MIGS in the $MX_{2}$ materials. For more details on the MIGS decay lengths in $MX_{2}$ monolayers please see \cite{szczesniak1}. Moreover, in Fig. \ref{fig02}, one can observe that the polarization becomes stronger via given transition metal Mo$\rightarrow$W and chalcogen S$\rightarrow$Se$\rightarrow$Te substitutions. This fact is in accordance with the growing SOC constant in the corresponding monolayers. Altogether, the above findings prove that the valley- and spin-polarization of MIGS is highly correlated, in agreement with the valley-spin locking effect. Most importantly however, the described valley-spin filtering behavior appears to be in qualitative agreement with the predictions made previously within other modeling studies \cite{ominato, hsieh, tahir, rostami, majidi}. In what follows, the presented results reinforce the important role of MIGS in the discussed filtering effect.

In addition to the already provided observations, the results depicted in Fig. \ref{fig02} also allow to discuss the valley-spin filtering effect with respect to the Fermi level position within the energy gap. Specifically, when the Fermi level is located at the midgap position, the valley-spin polarization is not impressive ($<50$\%). However, the situation changes when the Fermi level position is raised (BP$^{+}$) or lowered (BP$^{-}$) with respect to the BP energy. In details, the $P_{S}$ and $P_{V}$ parameters approach level of 75\% in the BP$^{+}$ case, and notably surpass it at the BP$^{-}$ energy. Note that this result is also valid for $L \sim 2-3$ nm, when the tunneling currents are still particularly effective. Consequently, the strongest valley-spin filtering effect is obtained when the Fermi level lies close to the valence band, what can be related to the substantial SOC splitting of the valence band. To this end, a closer investigation of the discussed results allows to observe that $P_{S}$ displays somewhat stronger dependence on $L$ than the $P_{V}$ parameter. This discrepancy is particularly visible at the BP$^{+}$ energy level, and appears because $P_{V}$ concerns about MIGS related to each other by the time reversal symmetry, which is not the case for the $P_{S}$ parameter (see Eq. (\ref{eq3}) and (\ref{eq4})).

\section{Summary and conclusions}

In summary, the conducted analysis shows that the mechanism of the valley-spin filtering effect in the electrode-terminated $MX_{2}$ monolayers can be largely explained by the concept of MIGS. This finding is argued to be of great importance to the future design of the $MX_{2}$-based filter devices that employ valley and spin degrees of freedom for information processing. In particular, the MIGS are shown to explicitly relate filtering processes of interest to the intrinsic electronic properties of the semiconducting channel. In this context, they stress the role of the electrode/$MX_{2}$ interfaces in the filtering process, but also suggest routes toward engineering efficient valley-spin tunnel devices ({\it e.g.} the tunnel field effect transitions).

The developed theoretical model also allows to draw general trends in tuning the filtering properties with respect to the out-of-plane magnetic field strength, the MIGS decay distance, as well as the position of the Fermi level within the energy gap. The obtained filtering characteristics appear to be in qualitative agreement with the available modeling studies \cite{ominato, hsieh, tahir, rostami, majidi}. In what follows, they should constitute relevant basis for further investigations, aimed at the enhancement of the valley and spin functionalities in a low-dimensional systems. In particular, it is suggested that the best valley and spin selectivity under external magnetic field can be achieved for the Te-based monolayers, when the Fermi level is located below the midgap position, according to the large SOC splitting of the valence band. 

However, it is crucial to remark here that the external magnetic field required for the valley and spin polarization is of the order of hundreds of Teslas, and the magnetic exchange fields should be considered as a practical polarization technique in this regard. In details, recent theoretical and experimental studies show that the large Zeeman splitting (up to 300 meV), in both Mo- and W-based $MX_{2}$ monolayers, can be generated by the the magnetic proximity coupling to the ferro- and anti-ferromagnetic substrates. Among others, the described effect is theoretically predicted in MoTe$_{2}$ on EuO \cite{qi, cortes} and WS$_{2}$ on MnO \cite{xu}, as well as experimentally observed in WSe$_{2}$ on EuS \cite{zhao}, WS$_{2}$ on EuS \cite{norden}. In this context, it is important to note that the mentioned magnetic proximity coupling generates effective magnetic field, experienced by the $MX_{2}$ materials, that is qualitatively well described by the presented here theoretical approach. For more technical details please see \cite{cortes}, where the on-site magnetic exchange (given here by the $\mathbf{H}_{\rm B}$ term in Eq. (\ref{eq2})) is shown to describe one of the major substrate effects. Nonetheless, the inclusion of an additional Rashba fields should be also of interest for the future investigations.

To this end, the presented analysis is expected to be of general benefit for the research on the valley-spin filtering effect in a 2D materials. In particular, due to the properties of the $MX_{2}$ monolayers, as well as the universal character of the MIGS, the reported results emerge as a case study of the valley-spin filtering in the entire class of the 2D hexagonal systems with a broken inversion symmetry. Simultaneously, the presented model holds the potential for explaining the valley-spin filtering in 2D materials, when the polarization is induced and controlled by the non-magnetic means, as the MIGS are directly related to the band structure of a filter material.

\section{Acknowledgements}

D. Szcz{\c e}{\' s}niak acknowledges financial support of this work and the related research activities by the Polish National Agency for Academic Exchange (NAWA) under Bekker's programme (project no. PPN/BEK/2018/1/00433/U/00001). S. Kais would like to acknowledge funding by the U.S. Department of Energy (Office of Basic Energy Sciences) under award number DE-SC0019215. The Authors would like to also acknowledge fruitful discussions with Dr. Z. Hu (Purdue University).

\bibliographystyle{apsrev}
\bibliography{bibliography}

\end{document}